\documentclass[fleqn,twoside,twocolumn,longbibliography]{revtex4-1} % Specifies the document class %,unsortedaddress
\pdfoutput=1
\usepackage{hyperref}
\usepackage{amssymb}
\usepackage{amsmath}
\usepackage{graphicx}
\begin{document}
\title[Projected Gross-Pitaevskii equation for ring-shaped Bose-Einstein condensates]%колонтитул
{Projected Gross-Pitaevskii equation for ring-shaped Bose-Einstein condensates}%
\author{O. O. Prikhodko}%1
\affiliation{Department of Physics, Taras Shevchenko National University of Kyiv, Volodymyrska Str. 64/13, Kyiv 01601, Ukraine}%адрес 1
\email{eap@univ.kiev.ua}%e-mail 1
\author{Y. M. Bidasyuk}%2
\affiliation{Physikalisch-Technische Bundesanstalt, Bundesallee 100, D-38116 Braunschweig, Germany}%адрес 2
\email{yuriy.bidasyuk@ptb.de}%e-mail 1

\begin{abstract}
We propose an alternative implementation of the Projected Gross-Pitaevskki equation adapted for numerical modeling of the atomic Bose-Einstein condensate trapped in a toroidally-shaped potential. We present an accurate and efficient scheme to evaluate the required matrix elements and calculate time evolution of the matter wave field. We analyze the stability and accuracy of the developed method for  equilibrium and nonequilibrium solutions in a ring-shaped trap with additional barrier potential corresponding to recent experimental realizations.
\end{abstract}


\maketitle

\section{Introduction}
Gross-Pitaevskii equation (GPE) is the most widely used mathematical tool to model atomic Bose-Einstein condensates (BEC) and their dynamics at zero temperature \cite{RevModPhys.71.463,pethick2008bose}. Various modifications have been proposed to extend the applicability of GPE for a non-perturbative treatment of finite temperature effects and non-equillibrium dynamics. Such methods are commonly termed classical-field (or $C$-field) methods. Most notable methods of this class are truncated Wigner approximation \cite{PhysRevLett.87.210404} and the
Projected Gross-Pitaevskii equation (PGPE) \cite{PhysRevLett.87.160402,PhysRevA.72.063608}. The latter one will be the main focus of the present work.
A wide range of physical problems addressed with PGPE and its modifications include in particular Bose-condensation and quasicondensation \cite{PhysRevA.68.053615,PhysRevA.93.063603,PhysRevA.87.063611},
dynamical generation~\cite{PhysRevA.88.063620} and decay \cite{PhysRevA.81.023630} of quantum vortices, dissipative bosonic Josephson effect \cite{Bidasyuk_2018}.

From the numerical point of view Projected Gross-Pitaevskii equation belongs to the class of pseudospectral methods. 
It relies on the reformulation of the GPE in the spectral basis of single-particle states and frequent transformations between coordinate and spectral representations are at the core of the numerical procedure.
Such an approach requires explicit knowledge of the basis states in order to efficiently transform the condensate wave function between the two representations. It is therefore quite natural, that
existing realizations of PGPE are based on the eigenstates of a three-dimensional harmonic oscillator potential 
\cite{PhysRevLett.87.160402,PhysRevA.72.063608,PhysRevE.78.026704}.
This limits the applicability of such realizations to the traps which can be well approximated by the harmonic oscillator and account for any non-harmonic part as a small perturbation.

In the present work we propose an extension of the PGPE formalism to describe Bose-Einstein condensates trapped in toroidally-shaped traps. 
While single particle states of a toroidal trap can not be obtained analytically, we show here that PGPE can be formulated equally well in terms of approximate eigenstates and produce physically relevant results.
We verify the accuracy and time stability of the developed approach and demonstrate that made approximations do not introduce significant errors.
The developed approach can be straightforwardly extended to include dynamical noise terms and implement the stochastic projected Gross-Pitaevskii equation (SPGPE). This will allow to model a dynamical evolution of finite-temperature toroidal condensates. 

\section{PGPE model for toroidal system}
We consider a system that is characterized by the mean field Gross-Pitaevskii Hamiltonian operator $H_\mathrm{GP}$ \cite{pethick2008bose,RevModPhys.71.463}:
\begin{multline}
H_\mathrm{GP}\, \psi(\mathbf{r},t) = \left[-\frac{\hbar^2\nabla^2}{2M} + V_\mathrm{trap}(\mathbf{r}) + \delta V(\mathbf{r},t)  \right. \\ \left. + g |\psi(\mathbf{r},t)|^2 \right] \psi(\mathbf{r},t).
\label{eq:gpham}
\end{multline}
with the nonlinear interaction parameter $g=4\pi\hbar^2 a/M$, where $a$ is the $s$-wave scattering length 
%of $^{23}\mathrm{Na}$ 
and $M$ is the atom mass.
The potential $V_\mathrm{trap}(\mathbf{r})$ is a cylindrically symmetric ring-shaped trap formed by a combination of a shifted harmonic potential in the radial direction and another harmonic potential in the vertical direction \cite{PhysRevA.91.033607,nature14}:
\begin{equation}
V_\mathrm{trap}(\mathbf{r}) = \frac{M}{2}\left[\omega_r^2(r-r_0)^2 + \omega_z^2 z^2\right],
\label{eq:ring_pot}
\end{equation}
where we use cylindrical coordinates $\mathbf{r} = \{r,\theta,z\}$,  $r = \sqrt{x^2+y^2} $. The additional time-dependent potential $\delta V(\mathbf{r},t)$ is considered as a (small) perturbation to the trap potential. It can represent, for example, a moving barrier as in experiments of Refs.~\cite{nature14,PhysRevA.95.021602}.

Classical field or $C$-field methods are based on the concept of splitting the many-particle system into highly occupied low-energy modes described by the coherent classical field $\psi(\mathbf{r},t)$ and sparsely occupied incoherent high-energy modes forming a thermal bath. 
Such splitting is 
conveniently represented in the basis of single-particle eigenstates $\phi_n$ of the trapping potential $V_\mathrm{trap}$
\begin{equation}
H_0 \phi_{\alpha} = \left[-\frac{\hbar^2\nabla^2}{2M} + V_\mathrm{trap}(\mathbf{r})\right]\phi_{\alpha} = E_{\alpha}\phi_{\alpha},
\label{eq:speq}
\end{equation}
where $\alpha$ represents a set of quantum numbers that characterize the single-particle eigenstates $\phi_{\alpha}$.
The classical field $\psi(\mathbf{r},t)$ is then a coherent superposition of these states with energies below the chosen cut-off energy $e_\mathrm{cut}$
\begin{align}\label{eq:oscrep}
&\psi (\mathbf{r},t) = \sum\limits_{\alpha \in C} c_{\alpha}(t) \phi_{\alpha}(\mathbf{r}), 
\\ &C = \{\alpha:E_{\alpha} \leq e_\mathrm{cut}\}. \notag
\end{align}
The choice of the cut-off energy may be a complicated problem for finite temperature calculations (see e.g. \cite{PhysRevA.86.033610,PhysRevA.98.023622}). In the case of zero temperature this parameter only determines the 
basis size and overall accuracy of the decomposition (\ref{eq:oscrep})

Unfortunately, for the ring-shaped potential (\ref{eq:ring_pot}) we can not solve the single-particle problem (\ref{eq:speq}) analytically. Instead we can  choose a basis that only approximately diagonalizes the Hamiltonian $H_0$.
We define the basis states for the ring-shaped system as
\begin{equation}
\phi_{\alpha} (r,\theta,z)  =  \frac{1}{\sqrt{2 \pi r}} \varphi_n^{(\omega_r)}(r-r_0) \varphi_m^{(\omega_z)}(z) e^{i l \theta},
\label{eq:ring_basis}
\end{equation}
where $\alpha$ contains now three quantum numbers $\alpha\rightarrow\{n,l,m\}$ and  $\varphi_n^{(\omega)}(x)$ are normalized eigenstates of a one-dimensional harmonic oscillator with frequency $\omega$:
\[
\varphi_n^{(\omega)}(x) = \sqrt{\frac{b}{2^n \sqrt{\pi} n!}} H_{n} \left( \frac{x}{b} \right) e^{-\frac{x^2}{2b^2}}, 
\]
where $b = \sqrt{\hbar/M\omega}$ is the characteristic oscillator length, $H_n$ is the Hermite polynomial of the order $n$.
This basis (\ref{eq:ring_basis}) is not orthonormalized due to its radial dependence. The approximate orthogonality can be ensured if $\sqrt{\hbar/M\omega_r} \ll r_0$ (see Appendix~\ref{app:orth} for more details). 
The Hamiltonian $H_0$ is also not fully diagonalized by the chosen basis, but rather takes the form
\begin{multline*}
\langle \phi_{n'l'm'} | H_0 |  \phi_{nlm} \rangle = \left[ (E^{(r)}_n + E^{(z)}_m) \delta_{nn'}  \right. \\ \left. + E^{(\theta)}_l I_{nn'} \right] \delta_{mm'} \delta_{ll'},
\end{multline*}
where 
\begin{align}
E^{(r)}_n &= \hbar\omega_r \left(n+\frac12 \right),\quad 
E^{(z)}_m = \hbar\omega_z \left(m+\frac12 \right), \notag\\
E^{(\theta)}_l &= \frac{\hbar^2}{2 M r_0^2} \left(l^2-\frac14\right), \notag\\
I_{nn'} &= \int\limits_{0}^{\infty} dr \frac{r_0^2}{r^2} \varphi_n^{(\omega_r)}(r-r_0) \varphi_{n'}^{(\omega_r)} (r-r_0). \label{eq:inn}
\end{align}

The matrix element $I_{nn'}$ formally diverges at $r\rightarrow 0$. 
It can still be meaningfully approximated if we restrict the integration to the region of finite support of the oscillator functions and use again the condition $\sqrt{\hbar/M\omega_r} \ll r_0$
(see Appendix~\ref{app:inn} for more details).
In this case $I_{nn'}$ is also close to identity matrix and we can approximately define the single-particle energy spectrum as
\begin{equation}\label{eq:spenergy}
E_{nml} = E^{(r)}_n + E^{(z)}_m + E^{(\theta)}_l.
\end{equation}
Using this approximate spectrum and chosen cut-off energy we define the $C$-region and truncate the basis~(\ref{eq:oscrep})
\[
C = \{n,m,l:E^{(r)}_n + E^{(z)}_m + E^{(\theta)}_l \leq e_\mathrm{cut}\},
\]
which also fixes the maximal value of each of the quantum numbers $n_{max}$, $m_{max}$, $l_{max}$.

The density of states which corresponds to the spectrum (\ref{eq:spenergy}) can be calculated analytically as follows
\begin{equation}\label{eq:dens_states0}
\rho(\epsilon) = \frac43 \frac{\sqrt{2M}r_0}{\hbar^3\omega_r\omega_z} \epsilon^{3/2}.
\end{equation}
More details on this derivation can be found in the Appendix~\ref{app:dos}.

The density of states can be also estimated in quasiclassical approximation
\begin{multline}
\rho_{qc}(\epsilon) = \int \frac{d\mathbf{r} d\mathbf{p}}{(2\pi\hbar)^3} \delta(\epsilon - E(\mathbf{r},\mathbf{p}))  \\ = \frac{M^{3/2}}{\sqrt2 \pi^2 \hbar^3}\int_{V \leq \epsilon} d\mathbf{r} \sqrt{\epsilon - V(\mathbf{r})},
\label{eq:dens_states1}
\end{multline}
where $E(\mathbf{r},\mathbf{p})$ is the energy of a classical particle in the potential $V(\mathbf{r}) = V_{trap}(\mathbf{r}) + \delta V(\mathbf{r})$.
The integral in (\ref{eq:dens_states1}) can be calculated analytically for a pure ring trap potential (\ref{eq:ring_pot}) and energies $\epsilon < M\omega_r^2 r_0^2/2$ producing the same expression as above. 
In general the closeness of the two estimates (\ref{eq:dens_states0}) and (\ref{eq:dens_states1}) shows how good the real spectrum
of Eq.~(\ref{eq:speq}) is reproduced by the approximate basis states (\ref{eq:ring_basis}). From the density of states (\ref{eq:dens_states0}) one may also see that the number of basis states in $C$-region (and consequently the numerical complexity of the calculations) grows with the cut-off as $N_C \sim e_{cut}^{5/2}$.

If we completely neglect the incoherent region (all single-particle states above the cut-off) then the classical field $\psi (\mathbf{r},t)$ will be a solution to the projected Gross-Pitaevskii equation (PGPE) \cite{PhysRevA.72.063608,PhysRevE.78.026704}:
\begin{equation}
i\hbar \frac{\partial \psi (\mathbf{r},t)}{\partial t} = \mathcal{P} H_\mathrm{GP} \psi(\mathbf{r},t)
\label{eq:spgpe}
\end{equation}
where $\mathcal{P}$ is a projection operator to the $C$-space.
\[
\mathcal{P} \psi (\mathbf{r},t) = \sum\limits_{\alpha\in C} \phi_{\alpha}(\mathbf{r}) \int d\mathbf{r}' \phi_{\alpha}^*(\mathbf{r}') \psi (\mathbf{r}',t).
\]

In the spectral basis the equation for expansion coefficients $c_{\alpha}$ reads

\begin{equation}
i\hbar \frac{d c_{\alpha}}{dt} = (E^{(r)}_n + E^{(z)}_m) c_{\alpha} + E^{(\theta)}_l D_{\alpha} + F_{\alpha}
\label{eq:spgpe2}
\end{equation}
where
\begin{equation}
D_{\alpha} = \int d\mathbf{r} \phi^*_{\alpha}(\mathbf{r}) \frac{r_0^2}{r^2} \psi (\mathbf{r},t),
\label{eq:spgpe_coef1}
\end{equation}
\begin{equation}
F_{\alpha} = \int d\mathbf{r} \phi^*_{\alpha}(\mathbf{r}) \left[ \delta V(\mathbf{r},t) + g |\psi(\mathbf{r},t)|^2 \right] \psi (\mathbf{r},t)  
\label{eq:spgpe_coef2}
\end{equation}

In order to numerically solve the Eq.~(\ref{eq:spgpe2}) we need an efficient and accurate way to transform the solution between the coordinate and spectral representations.
The integrals containing harmonic oscillator states can be accurately approximated by the Gauss-Hermite quadrature. The general form of the $N_Q$ point quadrature rule is
\[
\int_{-\infty}^{\infty} dx e^{-x^2} f(x) \approx \sum\limits_{j=1}^{N_Q} w_j f(x_j),
\]
where $x_j$ and $w_j$ are the quadrature points and weights. This quadrature rule is exact if $f(x)$ is a polynomial of a degree below $2N_Q-1$.
Transformation of any function $\psi(\mathbf{r})$ to the basis representation is then constructed as follows:
\begin{multline*}
c_{nlm} = \int d\mathbf{r} \phi^*_{nlm}(\mathbf{r}) \psi(\mathbf{r}) \\
=  \sum\limits_{jks} w^{(r)}_j w^{(z)}_s \delta\theta U_{jn}\, W^*_{kl}\, Y_{sm} \psi(r_j,\theta_k,z_s),
\end{multline*}
where we introduce the rescaled quadrature weights
\[
w^{(r)}_j = w_j b_{r} e^{(r_j-r_0)^2/b_r^2},\qquad w^{(z)}_s = w_s b_{z} e^{r_s^2/b_z^2},
\]
with
\[
b_r = \sqrt{\frac{\hbar}{M\omega_r}}, \qquad b_z = \sqrt{\frac{\hbar}{M\omega_z}}.
\]
Integration in the azimuthal direction is performed with a usual trapezoidal rule on a uniform grid with spacing $\delta\theta$.
The transformation matrices are defined as the basis states evaluated on the quadrature grid:
\[
U_{jn} = \varphi^{(\omega_r)}_n(r_j-r_0),\,\,\,\,
W_{kl} = e^{il\theta_k},\,\,\,\,
Y_{sm} = \varphi^{(\omega_z)}_m(z_s).
\]
The backwards transformation to the spatial representation is then performed as follows:
\[
\psi(r_j,\theta_k,z_s) = \sum\limits_{nml} U_{jn}\, W_{kl}\, Y_{sm} \,c_{nlm}.
\]

For more details on the transformations between coordinate and spectral representations and calculation of matrix elements we refer to Ref.~\cite{PhysRevE.78.026704}. It is worth noticing that in practical realizations the transformation with matrix $W_{kl}$ can be replaced with a Fast Fourier Transform for better performance. We however prefer to keep this transformation matrix here for clarity.

In order to perform a time evolution of the Eq.~(\ref{eq:spgpe2}) we build a computational scheme similar to the split-step Fourier transform (SSFT) method which is widely used for GPE modeling \cite{BAO2003318}. 
This method implements a time evolution operator $\exp(-iH_{GP}t/\hbar)$ to propagate the condensate wave function in time.
Adapting this scheme to PGPE (\ref{eq:spgpe2}) and using a second order Trotter decomposition for the time evolution operator a basic time evolution step $c_{nlm}(t) \rightarrow c_{nlm}(t+\delta t)$ can be outlined as the following sequence:
\begin{align*}
1\!: &\,\,\, c'_{nlm} =  \exp\left[ - \frac{i\delta t}{2\hbar} (E^{(r)}_n + E^{(z)}_m) \right] c_{nlm}(t), \\
2\!: &\,\,\, d_{jlm} =  \exp\left[ - \frac{i\delta t}{2\hbar} E^{(\theta)}_l \frac{r_0^2}{r_j^2} \right] \sum_{n} U_{jn} c'_{nlm},\\
3\!: &\,\,\, f_{jks} = \sum_{lm} W_{kl} Y_{sm} d_{jlm},\\
4\!: &\,\,\, f'_{jks} = \exp\left[ - \frac{i\delta t}{\hbar} \left(\delta V(r_j,\theta_k,z_s,t) + g |f_{jks}|^2 \right) \right] f_{jks}, \\
5\!: &\,\,\, d'_{jlm}  = \sum_{ks} w_s^{(z)} \delta\theta \, W^*_{kl} Y_{sm} f'_{jks},\\
6\!: &\,\,\, c''_{nlm} =  \sum_{j} w_j^{(r)} U_{jn} \exp\left[ - \frac{i\delta t}{2\hbar} E^{(\theta)}_l \frac{r_0^2}{r_j^2} \right] d'_{jlm}, \\
7\!: &\,\,\, c_{nlm}(t+\delta t) =  \exp\left[ - \frac{i\delta t}{2\hbar} (E^{(r)}_n + E^{(z)}_m) \right] c''_{nlm}. \\
\end{align*}
We note that in order to calculate the term which includes the integral $D_{\alpha}$ defined by Eq.~(\ref{eq:spgpe_coef1}) we need to perform a partial transformation and use coordinate representation in $r$ together with a spectral representation in $\theta$ and $z$.

\section{Numerical verification}

In order to test the developed numerical approach we model the 
toroidal trap of the experiment  \cite{nature14}. The parameters of the trap potential are then defined as follows: $\omega_r/2\pi=188~\mathrm{Hz}$, $\omega_z/2\pi=472~\mathrm{Hz}$, $r_0=19.5~\mathrm{\mu m}$. The total number of atoms in BEC is $N=4\cdot10^5$ and corresponding chemical potential is estimated as $\mu\approx 10\hbar\omega_r$. The barrier is approximated by a following potential, which for the purposes of present study we consider as time-independent: 
\[
\delta V (\mathbf{r}) = V_b \Theta(x) e^{-\frac{y^2}{2\lambda^2}},
\]
where $\Theta(x)$ is a Heaviside step function, $\lambda = 6 \mu$m is the $1/e^2$ half width of the barrier and we choose  the barrier height to match the value of the chemical potential $V_b=10\hbar\omega$.

\begin{figure}
	\vskip1mm
	\includegraphics[width=\linewidth]{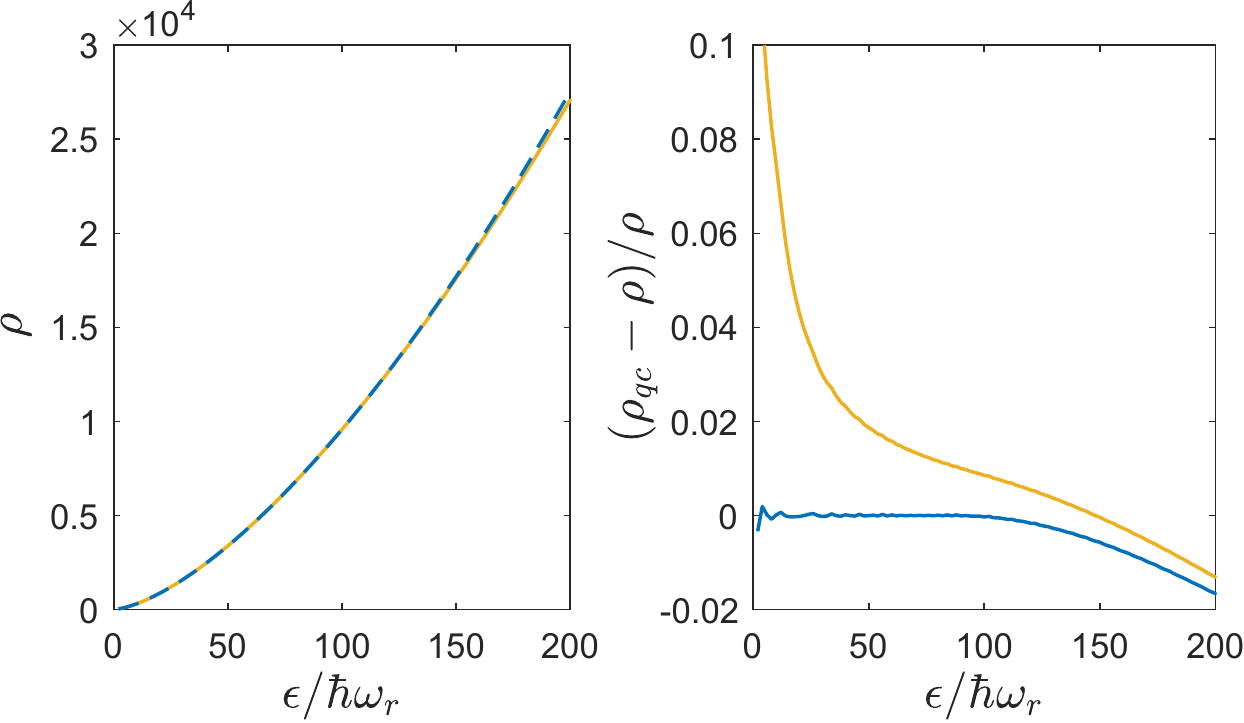}
	\vskip-3mm\caption{ Left panel: Density of states for the ring-shaped potential without a barrier from Eq.~(\ref{eq:dens_states0}) (dashed blue line) and  (\ref{eq:dens_states1}) (solid yellow line). Right panel: Blue (dark grey) line shows relative error of Eq.~(\ref{eq:dens_states0}) for a homogeneous ring, yellow (light grey) line is the same but for a ring with additional barrier potential.}
	\label{fig:densstates}
\end{figure}

The main requirement for the validity of our approach is $b_r \ll r_0$. For the trap parameters defined above we get $b_r/r_0 \approx 0.04$.
We first test the quality of our basis representation by evaluating the density of states and comparing it to the analytical expression (\ref{eq:dens_states0}). The result is shown in Fig.~\ref{fig:densstates}. It shows that the energy spectrum of a toroidal trap (without a barrier) is reproduced very accurately for energies up to $100\hbar\omega_r$. The discrepancy is expectedly higher when the barrier potential is taken into account. The relative error is however within 2\% in high-energy region which is very good for such a simple approximation and justifies the cut-off definition based on the the approximate spectrum (\ref{eq:spenergy}).

\begin{figure}
	\vskip1mm
	\includegraphics[width=\linewidth]{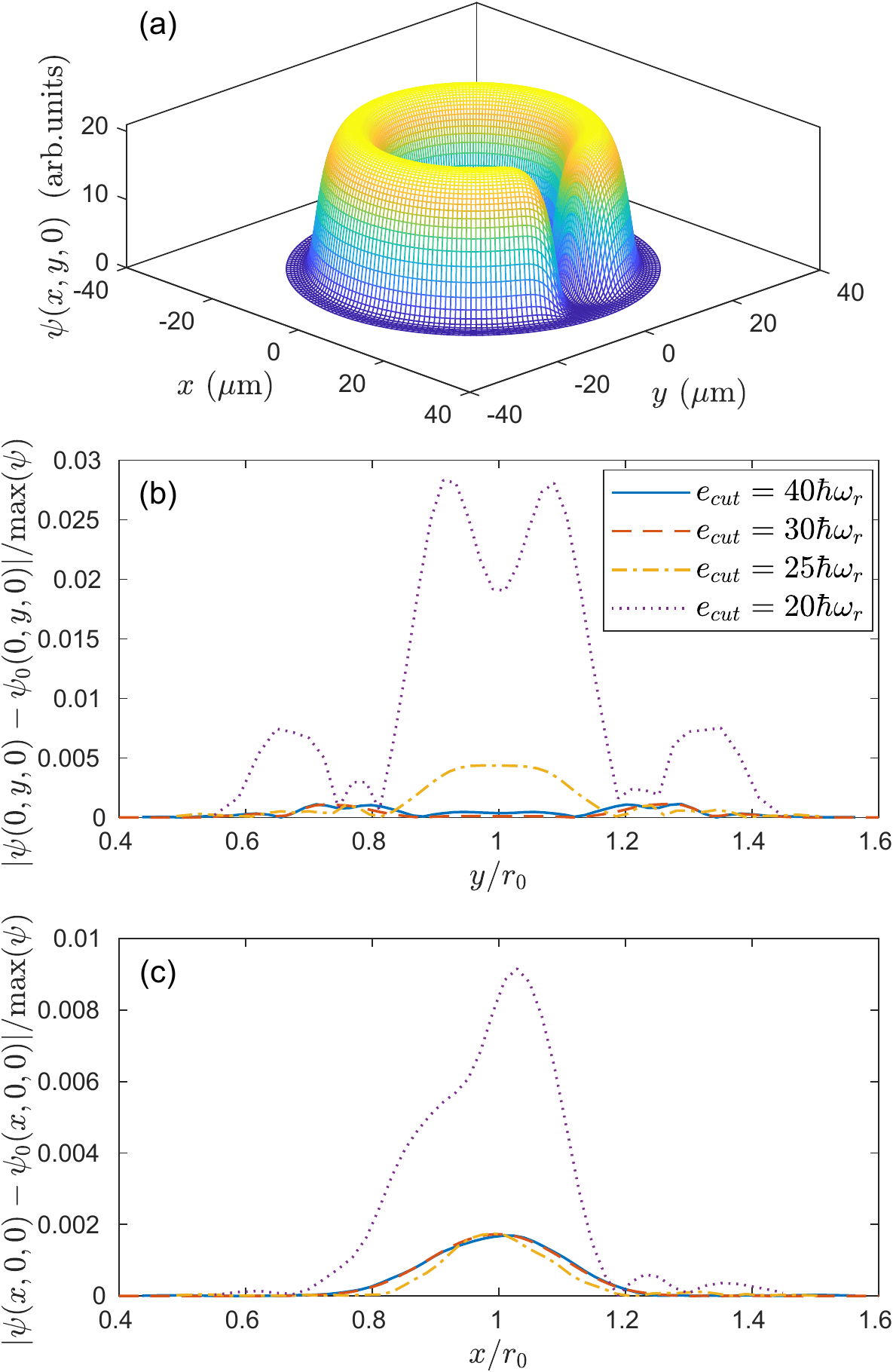}
	\vskip-3mm\caption{ (a) Ground state solution of PGPE on the quadrature points in the $z=0$ plane. (b) Relative error of the solution in coordinate space along the radial direction away from the barrier for four different values of the energy cut-off and consequently different basis sizes: $e_{cut}= 20\hbar\omega_r$ (7282 basis states), $e_{cut}= 25\hbar\omega_r$ (12576 basis states), $e_{cut}= 30\hbar\omega_r$ (19676 basis states), $e_{cut}= 40\hbar\omega_r$ (39970 basis states). (c) Same as (b) but along the barrier direction.}
	\label{fig:gs}
\end{figure}

Next, we calculate the ground state of the system with a barrier by propagating the PGPE (\ref{eq:spgpe2}) in the imaginary time. This is done for different values of $e_{cut}$ to see the effect of basis size on the accuracy of the calculated ground state. The results are shown at Fig.~\ref{fig:gs}. 
In order to estimate the error we compare the coordinate space representation of the obtained solutions to the solution of a three-dimensional GPE obtained on a very dense coordinate grid with the usual SSFT method. 
We see that for all chosen values of the cut-off energy our numerical procedure produces reasonable approximations of the condensate ground state. The error converges rapidly with increasing basis size and reaches saturation around $e_{cut} = 30\hbar\omega_r$. 
We conclude that this is the optimal cut-off energy for such system and use only this value for the rest of this section.
It is worth noticing that in realistic finite-temperature calculations the choice of the cut-off energy is a nontrivial problem and its definition is related to the temperature of the system \cite{PhysRevA.62.063609,PhysRevA.81.023630,Bidasyuk_2018}.
For the purposes of present feasibility study, which does not address any real finite-temperature processes, the cut-off value is considered only as a measure of the basis size and consequently the quality of spectral representation of the condensate wave function.

\begin{figure}
	\vskip1mm
	\includegraphics[width=\linewidth]{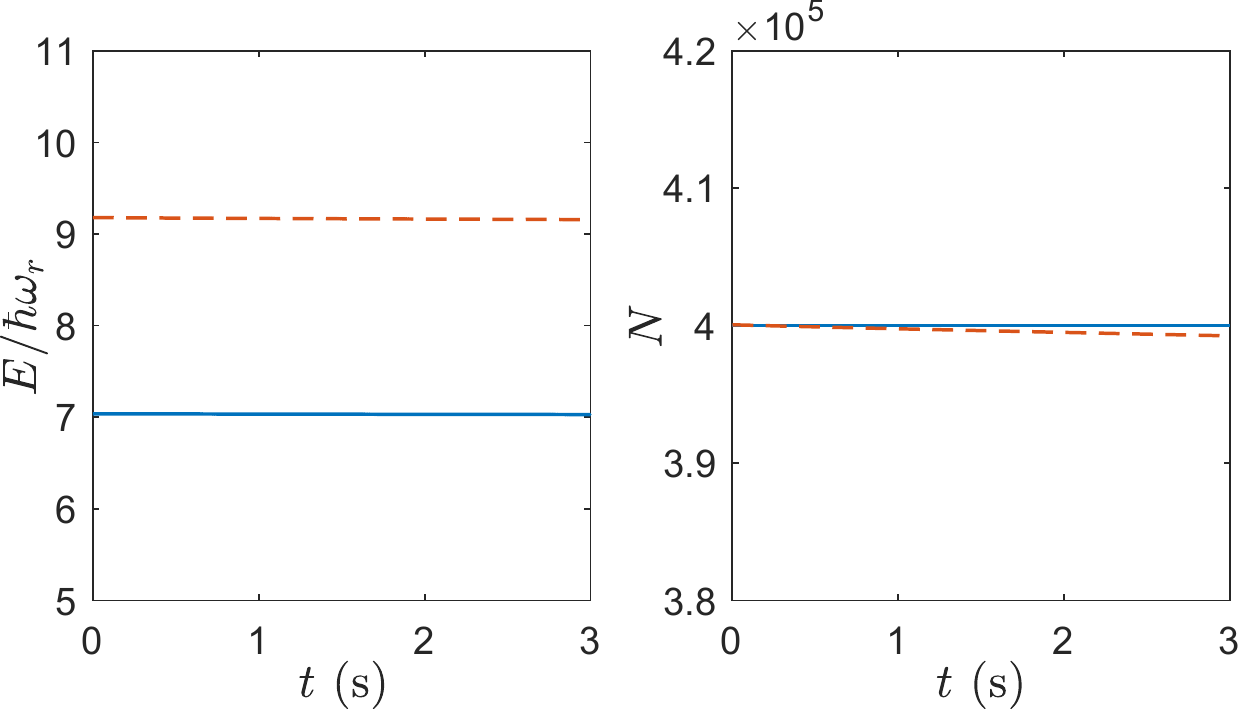}
	\vskip-3mm\caption{Energy per particle (Left panel) and total number of atoms (Right panel) monitored during the real time evolution of PGPE. The initial state for the evolution is chosen as equilibrium state (solid blue lines) or a non-equilibrium state (dashed red lines) with the same initial number of particles.}
	\label{fig:stab}
\end{figure}

While PGPE in general conserves the total energy and the normaization of the wave function (which is the total number of particles in the system) we can not prove that this conservation laws are preserved in the basis (\ref{eq:ring_basis}) which is only approximately orthogonal. This may lead to an accumulation of numerical errors and as a result to a drift of the conserved quantities. Such effects can be even stronger in the presence of the barrier potential as the single particle spectrum is shifted. We therefore check next that the energy and the atom number are reasonably conserved on a time scale of the experiment which is around 3 seconds in \cite{nature14}.
In order to prepare a non-equilibrium state we add to the stationary state a random complex noise uniformly distributed across all basis states. We then renormalize obtained state to obtain the state with the same number of atoms but with the higher energy then the ground state. Fig.~\ref{fig:stab} shows the evolution of the energy per particle and the number of particles in time for initial equilibrium and non-equilibrium states. The relative drift of these quantities on the time scale of the experiment is about 0.2\% for a non-equilibrium state. In the evolution of the stationary state no noticeable drift is observed.
Stability of the conserved quantities even for non-equilibrium states shows the applicability of the proposed time evolution scheme and overall consistency of the developed algorithm on physically relevant time scales.

\begin{figure}
	\vskip1mm
	\includegraphics[width=\linewidth]{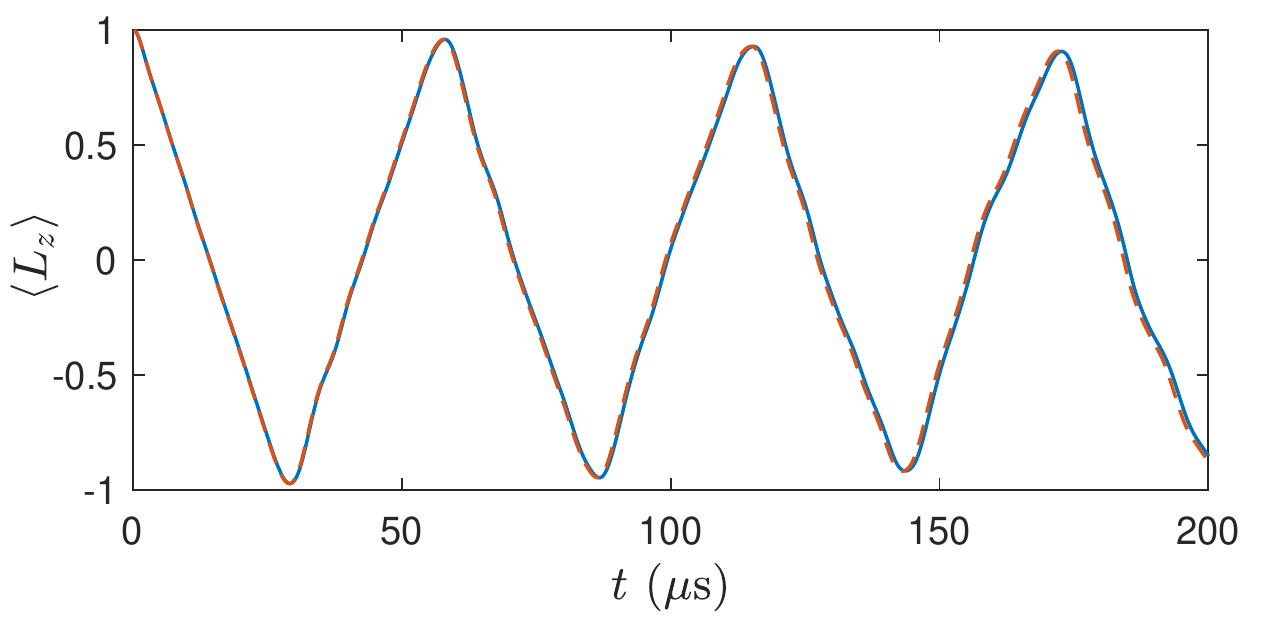}
	\vskip-3mm\caption{Evolution of the angular momentum projection $\langle L_z \rangle$ modeled with PGPE (solid blue line) and grid-based GPE (dashed red line).}
	\label{fig:lz}
\end{figure}

We perform the next test in order to further verify the accuracy of non-equilibrium dynamics reproduced by our evolution scheme. We prepare the initial state by adding a phase circulation to the stationary ground state introducing a single quantum of angular momentum to the system. Time evolution of such state effectively mimics the instability of a persistent current states in a ring with a barrier. As our equation does not contain any explicit dissipation mechanism such instability manifests as oscillations of the average angular momentum projection $\langle L_z \rangle$. Such unstable evolution was modeled with our evolution scheme of PGPE and with the grid-based GPE for comparison (see Fig.~\ref{fig:lz}). We see a nearly perfect match of the two results.
It is worth mentioning that the value of $\langle L_z \rangle$ can be calculated in the basis representation exactly as the basis states (\ref{eq:ring_basis}) are eigenfunctions of $L_z$ operator.

\section{Conclusions}

We have developed an implementation of a projected Gross-Pitaevskii equation adapted for Bose-Einstein condensates in toroidal traps. It is based on approximate eigenstates of a single-particle Hamiltonian which nevertheless closely reproduces the spectrum of the trap.

We have also proposed a time propagation scheme for PGPE which is similar to a well established split-step Fourier transform method. 
This scheme can be applied for both real and imaginary time evolution of PGPE. It was thoroughly tested and is shown to produce stable and accurate results.
Such fully explicit time evolution algorithm is straightforward to complement with time-dependent noise terms and realize a stochastic projected Gross-Pitaevskii equation. This will allow for modeling of various fintie-temperature processes in BEC which is the main application of PGPE models.
We therefore believe that the proposed method can be especially useful for modeling of the temperature-induced decay of persistent currents in BEC and will help to resolve existing discrepancies between theory and experiment \cite{nature14,PhysRevA.95.021602,PhysRevA.94.063642,PhysRevA.99.043613}. 

From the performance point of view, the advantage of PGPE is that it needs to be propagated on a relatively small prescribed basis, much smaller then the typical number of points in three-dimensional grid-based calculations. For the chosen value of the cut-off energy the basis size is about 20k states. If compared to grid-based calculations, the minimally acceptable three-dimensional grid size for the system under study can be estimated as $128\times128\times32$, which leads to more that 500k grid points. 
On the other hand, however, the effect of small basis size for PGPE is compensated by the additional computational cost of frequent transformations. 
Without performing a detailed performance study we only note that practical computational times were comparable for our implementations of PGPE and grid-based GPE.

%\vskip3mm \textit{Acknowledgement.}

\appendix

\section{Approximate orthogonality of the basis}\label{app:orth}

The basis (\ref{eq:ring_basis}) is only approximately orthonormalized due to its radial dependence. 
%Its radial part must be negligibly small at $r=0$ so that we can approximately consider
The overlap integral of two basis functions is
\begin{multline}
S_{\alpha\alpha'} = \iiint rdr\, d\theta\, dz \phi_{\alpha}(r,\theta,z) \phi_{\alpha'}(r,\theta,z) \\ = 
\delta_{ll'}\delta_{mm'}\int\limits_{0}^{\infty} dr \varphi_n^{(\omega_r)}(r-r_0) \varphi_{n'}^{(\omega_r)} (r-r_0)  \\
=\delta_{ll'}\delta_{mm'}\int\limits_{-r_0}^{\infty} dr \varphi_n^{(\omega_r)}(r) \varphi_{n'}^{(\omega_r)} (r).
% \\
%\approx \int\limits_{-\infty}^{\infty} dr \varphi_n^{(\omega_r)}(r) \varphi_{n'}^{(\omega_r)}(r) = \delta_{nn'}. 
\label{eq:r-ort}
\end{multline}
The oscillator functions have finite support defined by the classical turning points $R_n=\sqrt{\frac{2n\hbar}{M\omega_r}}$. Outside these points the function is exponentially small. Therefore if $r_0>R_n,R_{n'}$, then we can approximate the overlap integral as follows
\begin{multline}
S_{\alpha\alpha'} =  
\delta_{ll'}\delta_{mm'}\left[\delta_{nn'}-\int\limits_{-\infty}^{-r_0} dr \varphi_n^{(\omega_r)}(r) \varphi_{n'}^{(\omega_r)} (r) \right] \\ = \delta_{ll'}\delta_{mm'}\left[\delta_{nn'} + \mathcal{O}\left(\left(\frac{r_0}{b_r}\right)^{n+n'}e^{-(\frac{r_0}{b_r})^2}\right) \right]  \approx \delta_{\alpha\alpha'}, \label{eq:r-ort2}
\end{multline}
where $b_r=\sqrt{\hbar/M\omega_r} \ll r_0$ is a necessary requirement for approximate orthogonality.

\section{Matrix $I_{nn'}$ and the approximate spectrum}\label{app:inn}

Here we analyze the matrix elements $I_{nn'}$ defined by (\ref{eq:inn}) and show the validity of the approximate spectrum (\ref{eq:spenergy}).
More specifically, in order to define the cut-off energy we need to approximate the high-energy region of the spectrum. Therefore we are interested mainly in the behavior of $I_{nn'}$ for $n,n' \gg 1$. In this region the basis functions are rapidly oscillating and the integral (\ref{eq:inn}) can be approximated using stationary phase arguments \cite{PhysRevC.82.064603,BIDASYUK201360}:
\begin{multline}
I_{nn'} = \int\limits_{0}^{\infty} dr \frac{r_0^2}{r^2} \varphi_n^{(\omega_r)}(r-r_0) \varphi_{n'}^{(\omega_r)} (r-r_0) \\
\approx \delta_{nn'}\frac12\left[\frac{r_0^2}{(r_0+R_n)^2} + \frac{r_0^2}{(r_0-R_n)^2}\right] \\ \approx \delta_{nn'} \left[ 1+ 3\frac{R_n^2}{r_0^2} \right] , \label{eq:inn-app}
\end{multline}
where $R_n = \sqrt{\frac{2n\hbar}{M\omega_r}}$ are the classical turning points of the oscillator states, which are at the same time the points of stationary phase. The approximation is only valid if the integrand $r_0^2/r^2$ is a smooth continuous function and both points of stationary phase are within the integration region. This imposes additional restriction $R_n<r_0$.
With the result (\ref{eq:inn-app}) we get the following spectrum
\[
E_{nml} = E^{(r)}_n + E^{(z)}_m + E^{(\theta)}_l + 3\hbar\omega_r\frac{b_r^4}{r_0^4} n \left(l^2-\frac14\right),
\]
where $b_r=\sqrt{\hbar/M\omega_r}$. The condition $b_r \ll r_0$, which was imposed to ensure the orthogonality of the basis states, allows here to neglect the last term and justify the approximate spectrum (\ref{eq:spenergy}).

\section{Derivation of the density of states}\label{app:dos}
Here we show how the relations between density of states (\ref{eq:dens_states0}), the approximate spectrum (\ref{eq:spenergy}) and the quasiclassical integral (\ref{eq:dens_states1}). We start with the spectrum (\ref{eq:spenergy}):
\begin{multline*}
E_{nml}=\hbar\omega_r\left(n+\frac12\right) + \hbar\omega_z\left(m+\frac12\right) \\ + \frac{\hbar^2}{2Mr_0^2}\left(l^2-\frac14\right).
\end{multline*}
We are mostly interested in high-energy behavior of the spectrum. Therefore, to simplify the calculations we first shift the spectrum so that the ground state ($n=m=l=0$) has zero energy:
\[
\tilde E_{nml}=\hbar\omega_r n + \hbar\omega_z m + \frac{\hbar^2 l^2}{2Mr_0^2}.
\]
The number of states with energies $\tilde E < \epsilon$ is defined as the sum
\[
N(\epsilon) = \sum\limits_{\tilde E_{nml}<\epsilon} 1.
\]
The simplest way to calculate this sum is to consider $n$, $m$ and $l$ as continuous variables and convert it to the integral
\[
N(\epsilon) = \iiint\limits_{\tilde E_{nml}<\epsilon} dn\,dm\,dl.
\]
This integral yields
\begin{equation}\label{eq:nstates0}
N(\epsilon) = \frac8{15} \frac{\sqrt{2M}r_0}{\hbar^3\omega_r\omega_z} \epsilon^{5/2}.
\end{equation}
The density of states is then calculated as the derivative of the above expression:
\begin{equation}\label{eq:dos-app0}
\rho(\epsilon) = \frac{dN(\epsilon)}{d\epsilon} = \frac43 \frac{\sqrt{2M}r_0}{\hbar^3\omega_r\omega_z} \epsilon^{3/2}.
\end{equation}

Another approach to calculate the density of states is based on the quasiclassical approximation. The energy of the classical particle in the potential $V(\mathbf{r})=V_{trap}(\mathbf{r})$ is 
\[
E(\mathbf{r},\mathbf{p}) = \frac{p^2}{2M} + V(\mathbf{r}) = \frac{p^2}{2M} + \frac{M}{2}\left[\omega_r^2(r-r_0)^2 + \omega_z^2 z^2\right]
\]
The density of states is then defined by the following integral:
\begin{multline}
\rho_{qc}(\epsilon) = \int \frac{d\mathbf{r} d\mathbf{p}}{(2\pi\hbar)^3} \delta(\epsilon - E(\mathbf{r},\mathbf{p}))  \\ = \frac{M}{\pi^2\hbar^3} \int\limits_{V \leq \epsilon} d\mathbf{r} \int dp\,p^2 \delta \left(p^2-2M[\epsilon-V(\mathbf{r})]\right)
\\ = \frac{M^{3/2}}{\sqrt2 \pi^2 \hbar^3}\int\limits_{V \leq \epsilon} d\mathbf{r} \sqrt{\epsilon - V(\mathbf{r})} \\ 
= \frac{M^{3/2}}{\sqrt2 \pi^2 \hbar^3}\int\limits_{V \leq \epsilon} d\mathbf{r} \sqrt{\epsilon - M/2\left[\omega_r^2(r-r_0)^2 + \omega_z^2 z^2\right]} \\
= \frac{2 \sqrt{2M} r_0 \epsilon^{3/2}}{ \pi \hbar^3 \omega_r\omega_z}\int\limits_{\tilde r^2 + \tilde z^2 \leq 1} d\tilde r d\tilde z \sqrt{1 - \tilde r^2 - \tilde z^2} \\
= \frac{2 \sqrt{2M} r_0 \epsilon^{3/2}}{ \pi \hbar^3 \omega_r\omega_z} \frac23 \pi = \frac43 \frac{\sqrt{2M}r_0}{\hbar^3\omega_r\omega_z} \epsilon^{3/2},
\label{eq:dens_states1-der}
\end{multline}
where we have used the condition $\epsilon < M\omega_r^2 r_0^2/2$. In this way we have obtained the density of states which is the same as Eq.~(\ref{eq:dos-app0}). It is worth noticing that two derivations are based on rather different set of approximations.

\bibliography{refs}

\begin{thebibliography}{23}%
\makeatletter
\providecommand \@ifxundefined [1]{%
 \@ifx{#1\undefined}
}%
\providecommand \@ifnum [1]{%
 \ifnum #1\expandafter \@firstoftwo
 \else \expandafter \@secondoftwo
 \fi
}%
\providecommand \@ifx [1]{%
 \ifx #1\expandafter \@firstoftwo
 \else \expandafter \@secondoftwo
 \fi
}%
\providecommand \natexlab [1]{#1}%
\providecommand \enquote  [1]{``#1''}%
\providecommand \bibnamefont  [1]{#1}%
\providecommand \bibfnamefont [1]{#1}%
\providecommand \citenamefont [1]{#1}%
\providecommand \href@noop [0]{\@secondoftwo}%
\providecommand \href [0]{\begingroup \@sanitize@url \@href}%
\providecommand \@href[1]{\@@startlink{#1}\@@href}%
\providecommand \@@href[1]{\endgroup#1\@@endlink}%
\providecommand \@sanitize@url [0]{\catcode `\\12\catcode `\$12\catcode
  `\&12\catcode `\#12\catcode `\^12\catcode `\_12\catcode `\%12\relax}%
\providecommand \@@startlink[1]{}%
\providecommand \@@endlink[0]{}%
\providecommand \url  [0]{\begingroup\@sanitize@url \@url }%
\providecommand \@url [1]{\endgroup\@href {#1}{\urlprefix }}%
\providecommand \urlprefix  [0]{URL }%
\providecommand \Eprint [0]{\href }%
\providecommand \doibase [0]{http://dx.doi.org/}%
\providecommand \selectlanguage [0]{\@gobble}%
\providecommand \bibinfo  [0]{\@secondoftwo}%
\providecommand \bibfield  [0]{\@secondoftwo}%
\providecommand \translation [1]{[#1]}%
\providecommand \BibitemOpen [0]{}%
\providecommand \bibitemStop [0]{}%
\providecommand \bibitemNoStop [0]{.\EOS\space}%
\providecommand \EOS [0]{\spacefactor3000\relax}%
\providecommand \BibitemShut  [1]{\csname bibitem#1\endcsname}%
\let\auto@bib@innerbib\@empty
%</preamble>
\bibitem [{\citenamefont {Dalfovo}\ \emph {et~al.}(1999)\citenamefont
  {Dalfovo}, \citenamefont {Giorgini}, \citenamefont {Pitaevskii},\ and\
  \citenamefont {Stringari}}]{RevModPhys.71.463}%
  \BibitemOpen
  \bibfield  {author} {\bibinfo {author} {\bibfnamefont {Franco}\ \bibnamefont
  {Dalfovo}}, \bibinfo {author} {\bibfnamefont {Stefano}\ \bibnamefont
  {Giorgini}}, \bibinfo {author} {\bibfnamefont {Lev~P.}\ \bibnamefont
  {Pitaevskii}}, \ and\ \bibinfo {author} {\bibfnamefont {Sandro}\ \bibnamefont
  {Stringari}},\ }\bibfield  {title} {\enquote {\bibinfo {title} {Theory of
  bose-einstein condensation in trapped gases},}\ }\href {\doibase
  10.1103/RevModPhys.71.463} {\bibfield  {journal} {\bibinfo  {journal} {Rev.
  Mod. Phys.}\ }\textbf {\bibinfo {volume} {71}},\ \bibinfo {pages} {463--512}
  (\bibinfo {year} {1999})}\BibitemShut {NoStop}%
\bibitem [{\citenamefont {Pethick}\ and\ \citenamefont
  {Smith}(2008)}]{pethick2008bose}%
  \BibitemOpen
  \bibfield  {author} {\bibinfo {author} {\bibfnamefont {Christopher~J}\
  \bibnamefont {Pethick}}\ and\ \bibinfo {author} {\bibfnamefont {Henrik}\
  \bibnamefont {Smith}},\ }\href {\doibase 10.1017/CBO9780511802850} {\emph
  {\bibinfo {title} {Bose--Einstein condensation in dilute gases}}}\ (\bibinfo
  {publisher} {Cambridge university press},\ \bibinfo {year}
  {2008})\BibitemShut {NoStop}%
\bibitem [{\citenamefont {Sinatra}\ \emph {et~al.}(2001)\citenamefont
  {Sinatra}, \citenamefont {Lobo},\ and\ \citenamefont
  {Castin}}]{PhysRevLett.87.210404}%
  \BibitemOpen
  \bibfield  {author} {\bibinfo {author} {\bibfnamefont {Alice}\ \bibnamefont
  {Sinatra}}, \bibinfo {author} {\bibfnamefont {Carlos}\ \bibnamefont {Lobo}},
  \ and\ \bibinfo {author} {\bibfnamefont {Yvan}\ \bibnamefont {Castin}},\
  }\bibfield  {title} {\enquote {\bibinfo {title} {Classical-field method for
  time dependent bose-einstein condensed gases},}\ }\href {\doibase
  10.1103/PhysRevLett.87.210404} {\bibfield  {journal} {\bibinfo  {journal}
  {Phys. Rev. Lett.}\ }\textbf {\bibinfo {volume} {87}},\ \bibinfo {pages}
  {210404} (\bibinfo {year} {2001})}\BibitemShut {NoStop}%
\bibitem [{\citenamefont {Davis}\ \emph {et~al.}(2001)\citenamefont {Davis},
  \citenamefont {Morgan},\ and\ \citenamefont
  {Burnett}}]{PhysRevLett.87.160402}%
  \BibitemOpen
  \bibfield  {author} {\bibinfo {author} {\bibfnamefont {M.~J.}\ \bibnamefont
  {Davis}}, \bibinfo {author} {\bibfnamefont {S.~A.}\ \bibnamefont {Morgan}}, \
  and\ \bibinfo {author} {\bibfnamefont {K.}~\bibnamefont {Burnett}},\
  }\bibfield  {title} {\enquote {\bibinfo {title} {Simulations of bose fields
  at finite temperature},}\ }\href {\doibase 10.1103/PhysRevLett.87.160402}
  {\bibfield  {journal} {\bibinfo  {journal} {Phys. Rev. Lett.}\ }\textbf
  {\bibinfo {volume} {87}},\ \bibinfo {pages} {160402} (\bibinfo {year}
  {2001})}\BibitemShut {NoStop}%
\bibitem [{\citenamefont {Blakie}\ and\ \citenamefont
  {Davis}(2005)}]{PhysRevA.72.063608}%
  \BibitemOpen
  \bibfield  {author} {\bibinfo {author} {\bibfnamefont {P.~Blair}\
  \bibnamefont {Blakie}}\ and\ \bibinfo {author} {\bibfnamefont {Matthew~J.}\
  \bibnamefont {Davis}},\ }\bibfield  {title} {\enquote {\bibinfo {title}
  {Projected gross-pitaevskii equation for harmonically confined bose gases at
  finite temperature},}\ }\href {\doibase 10.1103/PhysRevA.72.063608}
  {\bibfield  {journal} {\bibinfo  {journal} {Phys. Rev. A}\ }\textbf {\bibinfo
  {volume} {72}},\ \bibinfo {pages} {063608} (\bibinfo {year}
  {2005})}\BibitemShut {NoStop}%
\bibitem [{\citenamefont {Davis}\ and\ \citenamefont
  {Morgan}(2003)}]{PhysRevA.68.053615}%
  \BibitemOpen
  \bibfield  {author} {\bibinfo {author} {\bibfnamefont {M.~J.}\ \bibnamefont
  {Davis}}\ and\ \bibinfo {author} {\bibfnamefont {S.~A.}\ \bibnamefont
  {Morgan}},\ }\bibfield  {title} {\enquote {\bibinfo {title} {Microcanonical
  temperature for a classical field: Application to bose-einstein
  condensation},}\ }\href {\doibase 10.1103/PhysRevA.68.053615} {\bibfield
  {journal} {\bibinfo  {journal} {Phys. Rev. A}\ }\textbf {\bibinfo {volume}
  {68}},\ \bibinfo {pages} {053615} (\bibinfo {year} {2003})}\BibitemShut
  {NoStop}%
\bibitem [{\citenamefont {Rooney}\ \emph {et~al.}(2016)\citenamefont {Rooney},
  \citenamefont {Allen}, \citenamefont {Z\"ulicke}, \citenamefont {Proukakis},\
  and\ \citenamefont {Bradley}}]{PhysRevA.93.063603}%
  \BibitemOpen
  \bibfield  {author} {\bibinfo {author} {\bibfnamefont {S.~J.}\ \bibnamefont
  {Rooney}}, \bibinfo {author} {\bibfnamefont {A.~J.}\ \bibnamefont {Allen}},
  \bibinfo {author} {\bibfnamefont {U.}~\bibnamefont {Z\"ulicke}}, \bibinfo
  {author} {\bibfnamefont {N.~P.}\ \bibnamefont {Proukakis}}, \ and\ \bibinfo
  {author} {\bibfnamefont {A.~S.}\ \bibnamefont {Bradley}},\ }\bibfield
  {title} {\enquote {\bibinfo {title} {Reservoir interactions of a vortex in a
  trapped three-dimensional bose-einstein condensate},}\ }\href {\doibase
  10.1103/PhysRevA.93.063603} {\bibfield  {journal} {\bibinfo  {journal} {Phys.
  Rev. A}\ }\textbf {\bibinfo {volume} {93}},\ \bibinfo {pages} {063603}
  (\bibinfo {year} {2016})}\BibitemShut {NoStop}%
\bibitem [{\citenamefont {Garrett}\ \emph {et~al.}(2013)\citenamefont
  {Garrett}, \citenamefont {Wright},\ and\ \citenamefont
  {Davis}}]{PhysRevA.87.063611}%
  \BibitemOpen
  \bibfield  {author} {\bibinfo {author} {\bibfnamefont {Michael~C.}\
  \bibnamefont {Garrett}}, \bibinfo {author} {\bibfnamefont {Tod~M.}\
  \bibnamefont {Wright}}, \ and\ \bibinfo {author} {\bibfnamefont {Matthew~J.}\
  \bibnamefont {Davis}},\ }\bibfield  {title} {\enquote {\bibinfo {title}
  {Condensation and quasicondensation in an elongated three-dimensional bose
  gas},}\ }\href {\doibase 10.1103/PhysRevA.87.063611} {\bibfield  {journal}
  {\bibinfo  {journal} {Phys. Rev. A}\ }\textbf {\bibinfo {volume} {87}},\
  \bibinfo {pages} {063611} (\bibinfo {year} {2013})}\BibitemShut {NoStop}%
\bibitem [{\citenamefont {Rooney}\ \emph {et~al.}(2013)\citenamefont {Rooney},
  \citenamefont {Neely}, \citenamefont {Anderson},\ and\ \citenamefont
  {Bradley}}]{PhysRevA.88.063620}%
  \BibitemOpen
  \bibfield  {author} {\bibinfo {author} {\bibfnamefont {S.~J.}\ \bibnamefont
  {Rooney}}, \bibinfo {author} {\bibfnamefont {T.~W.}\ \bibnamefont {Neely}},
  \bibinfo {author} {\bibfnamefont {B.~P.}\ \bibnamefont {Anderson}}, \ and\
  \bibinfo {author} {\bibfnamefont {A.~S.}\ \bibnamefont {Bradley}},\
  }\bibfield  {title} {\enquote {\bibinfo {title} {Persistent-current formation
  in a high-temperature bose-einstein condensate: An experimental test for
  classical-field theory},}\ }\href {\doibase 10.1103/PhysRevA.88.063620}
  {\bibfield  {journal} {\bibinfo  {journal} {Phys. Rev. A}\ }\textbf {\bibinfo
  {volume} {88}},\ \bibinfo {pages} {063620} (\bibinfo {year}
  {2013})}\BibitemShut {NoStop}%
\bibitem [{\citenamefont {Rooney}\ \emph {et~al.}(2010)\citenamefont {Rooney},
  \citenamefont {Bradley},\ and\ \citenamefont {Blakie}}]{PhysRevA.81.023630}%
  \BibitemOpen
  \bibfield  {author} {\bibinfo {author} {\bibfnamefont {S.~J.}\ \bibnamefont
  {Rooney}}, \bibinfo {author} {\bibfnamefont {A.~S.}\ \bibnamefont {Bradley}},
  \ and\ \bibinfo {author} {\bibfnamefont {P.~B.}\ \bibnamefont {Blakie}},\
  }\bibfield  {title} {\enquote {\bibinfo {title} {{Decay of a quantum vortex:
  Test of nonequilibrium theories for warm Bose-Einstein condensates}},}\
  }\href {\doibase 10.1103/PhysRevA.81.023630} {\bibfield  {journal} {\bibinfo
  {journal} {Phys. Rev. A}\ }\textbf {\bibinfo {volume} {81}},\ \bibinfo
  {pages} {023630} (\bibinfo {year} {2010})}\BibitemShut {NoStop}%
\bibitem [{\citenamefont {Bidasyuk}\ \emph {et~al.}(2018)\citenamefont
  {Bidasyuk}, \citenamefont {Weyrauch}, \citenamefont {Momme},\ and\
  \citenamefont {Prikhodko}}]{Bidasyuk_2018}%
  \BibitemOpen
  \bibfield  {author} {\bibinfo {author} {\bibfnamefont {Y~M}\ \bibnamefont
  {Bidasyuk}}, \bibinfo {author} {\bibfnamefont {M}~\bibnamefont {Weyrauch}},
  \bibinfo {author} {\bibfnamefont {M}~\bibnamefont {Momme}}, \ and\ \bibinfo
  {author} {\bibfnamefont {O~O}\ \bibnamefont {Prikhodko}},\ }\bibfield
  {title} {\enquote {\bibinfo {title} {Finite-temperature dynamics of a bosonic
  josephson junction},}\ }\href {\doibase 10.1088/1361-6455/aae022} {\bibfield
  {journal} {\bibinfo  {journal} {Journal of Physics B: Atomic, Molecular and
  Optical Physics}\ }\textbf {\bibinfo {volume} {51}},\ \bibinfo {pages}
  {205301} (\bibinfo {year} {2018})}\BibitemShut {NoStop}%
\bibitem [{\citenamefont {Blakie}(2008)}]{PhysRevE.78.026704}%
  \BibitemOpen
  \bibfield  {author} {\bibinfo {author} {\bibfnamefont {P.~Blair}\
  \bibnamefont {Blakie}},\ }\bibfield  {title} {\enquote {\bibinfo {title}
  {Numerical method for evolving the projected gross-pitaevskii equation},}\
  }\href {\doibase 10.1103/PhysRevE.78.026704} {\bibfield  {journal} {\bibinfo
  {journal} {Phys. Rev. E}\ }\textbf {\bibinfo {volume} {78}},\ \bibinfo
  {pages} {026704} (\bibinfo {year} {2008})}\BibitemShut {NoStop}%
\bibitem [{\citenamefont {Yakimenko}\ \emph {et~al.}(2015)\citenamefont
  {Yakimenko}, \citenamefont {Bidasyuk}, \citenamefont {Weyrauch},
  \citenamefont {Kuriatnikov},\ and\ \citenamefont
  {Vilchinskii}}]{PhysRevA.91.033607}%
  \BibitemOpen
  \bibfield  {author} {\bibinfo {author} {\bibfnamefont {A.~I.}\ \bibnamefont
  {Yakimenko}}, \bibinfo {author} {\bibfnamefont {Y.~M.}\ \bibnamefont
  {Bidasyuk}}, \bibinfo {author} {\bibfnamefont {M.}~\bibnamefont {Weyrauch}},
  \bibinfo {author} {\bibfnamefont {Y.~I.}\ \bibnamefont {Kuriatnikov}}, \ and\
  \bibinfo {author} {\bibfnamefont {S.~I.}\ \bibnamefont {Vilchinskii}},\
  }\bibfield  {title} {\enquote {\bibinfo {title} {Vortices in a toroidal
  bose-einstein condensate with a rotating weak link},}\ }\href {\doibase
  10.1103/PhysRevA.91.033607} {\bibfield  {journal} {\bibinfo  {journal} {Phys.
  Rev. A}\ }\textbf {\bibinfo {volume} {91}},\ \bibinfo {pages} {033607}
  (\bibinfo {year} {2015})}\BibitemShut {NoStop}%
\bibitem [{\citenamefont {{Eckel}}\ \emph {et~al.}(2014)\citenamefont
  {{Eckel}}, \citenamefont {{Lee}}, \citenamefont {{Jendrzejewski}},
  \citenamefont {{Murray}}, \citenamefont {{Clark}}, \citenamefont {{Lobb}},
  \citenamefont {{Phillips}}, \citenamefont {{Edwards}},\ and\ \citenamefont
  {{Campbell}}}]{nature14}%
  \BibitemOpen
  \bibfield  {author} {\bibinfo {author} {\bibfnamefont {S.}~\bibnamefont
  {{Eckel}}}, \bibinfo {author} {\bibfnamefont {J.~G.}\ \bibnamefont {{Lee}}},
  \bibinfo {author} {\bibfnamefont {F.}~\bibnamefont {{Jendrzejewski}}},
  \bibinfo {author} {\bibfnamefont {N.}~\bibnamefont {{Murray}}}, \bibinfo
  {author} {\bibfnamefont {C.~W.}\ \bibnamefont {{Clark}}}, \bibinfo {author}
  {\bibfnamefont {C.~J.}\ \bibnamefont {{Lobb}}}, \bibinfo {author}
  {\bibfnamefont {W.~D.}\ \bibnamefont {{Phillips}}}, \bibinfo {author}
  {\bibfnamefont {M.}~\bibnamefont {{Edwards}}}, \ and\ \bibinfo {author}
  {\bibfnamefont {G.~K.}\ \bibnamefont {{Campbell}}},\ }\bibfield  {title}
  {\enquote {\bibinfo {title} {{Quantized hysteresis in a superfluid atomtronic
  circuit}},}\ }\href {\doibase 10.1038/nature12958} {\bibfield  {journal}
  {\bibinfo  {journal} {Nature}\ }\textbf {\bibinfo {volume} {506}},\ \bibinfo
  {pages} {200--203} (\bibinfo {year} {2014})}\BibitemShut {NoStop}%
\bibitem [{\citenamefont {Kumar}\ \emph {et~al.}(2017)\citenamefont {Kumar},
  \citenamefont {Eckel}, \citenamefont {Jendrzejewski},\ and\ \citenamefont
  {Campbell}}]{PhysRevA.95.021602}%
  \BibitemOpen
  \bibfield  {author} {\bibinfo {author} {\bibfnamefont {A.}~\bibnamefont
  {Kumar}}, \bibinfo {author} {\bibfnamefont {S.}~\bibnamefont {Eckel}},
  \bibinfo {author} {\bibfnamefont {F.}~\bibnamefont {Jendrzejewski}}, \ and\
  \bibinfo {author} {\bibfnamefont {G.~K.}\ \bibnamefont {Campbell}},\
  }\bibfield  {title} {\enquote {\bibinfo {title} {Temperature-induced decay of
  persistent currents in a superfluid ultracold gas},}\ }\href {\doibase
  10.1103/PhysRevA.95.021602} {\bibfield  {journal} {\bibinfo  {journal} {Phys.
  Rev. A}\ }\textbf {\bibinfo {volume} {95}},\ \bibinfo {pages} {021602}
  (\bibinfo {year} {2017})}\BibitemShut {NoStop}%
\bibitem [{\citenamefont {Cockburn}\ and\ \citenamefont
  {Proukakis}(2012)}]{PhysRevA.86.033610}%
  \BibitemOpen
  \bibfield  {author} {\bibinfo {author} {\bibfnamefont {S.~P.}\ \bibnamefont
  {Cockburn}}\ and\ \bibinfo {author} {\bibfnamefont {N.~P.}\ \bibnamefont
  {Proukakis}},\ }\bibfield  {title} {\enquote {\bibinfo {title} {Ab initio
  methods for finite-temperature two-dimensional bose gases},}\ }\href
  {\doibase 10.1103/PhysRevA.86.033610} {\bibfield  {journal} {\bibinfo
  {journal} {Phys. Rev. A}\ }\textbf {\bibinfo {volume} {86}},\ \bibinfo
  {pages} {033610} (\bibinfo {year} {2012})}\BibitemShut {NoStop}%
\bibitem [{\citenamefont {Pietraszewicz}\ and\ \citenamefont
  {Deuar}(2018)}]{PhysRevA.98.023622}%
  \BibitemOpen
  \bibfield  {author} {\bibinfo {author} {\bibfnamefont {J.}~\bibnamefont
  {Pietraszewicz}}\ and\ \bibinfo {author} {\bibfnamefont {P.}~\bibnamefont
  {Deuar}},\ }\bibfield  {title} {\enquote {\bibinfo {title} {Classical fields
  in the one-dimensional bose gas: Applicability and determination of the
  optimal cutoff},}\ }\href {\doibase 10.1103/PhysRevA.98.023622} {\bibfield
  {journal} {\bibinfo  {journal} {Phys. Rev. A}\ }\textbf {\bibinfo {volume}
  {98}},\ \bibinfo {pages} {023622} (\bibinfo {year} {2018})}\BibitemShut
  {NoStop}%
\bibitem [{\citenamefont {Bao}\ \emph {et~al.}(2003)\citenamefont {Bao},
  \citenamefont {Jaksch},\ and\ \citenamefont {Markowich}}]{BAO2003318}%
  \BibitemOpen
  \bibfield  {author} {\bibinfo {author} {\bibfnamefont {Weizhu}\ \bibnamefont
  {Bao}}, \bibinfo {author} {\bibfnamefont {Dieter}\ \bibnamefont {Jaksch}}, \
  and\ \bibinfo {author} {\bibfnamefont {Peter~A.}\ \bibnamefont {Markowich}},\
  }\bibfield  {title} {\enquote {\bibinfo {title} {Numerical solution of the
  gross–pitaevskii equation for bose–einstein condensation},}\ }\href
  {\doibase https://doi.org/10.1016/S0021-9991(03)00102-5} {\bibfield
  {journal} {\bibinfo  {journal} {Journal of Computational Physics}\ }\textbf
  {\bibinfo {volume} {187}},\ \bibinfo {pages} {318 -- 342} (\bibinfo {year}
  {2003})}\BibitemShut {NoStop}%
\bibitem [{\citenamefont {Bijlsma}\ \emph {et~al.}(2000)\citenamefont
  {Bijlsma}, \citenamefont {Zaremba},\ and\ \citenamefont
  {Stoof}}]{PhysRevA.62.063609}%
  \BibitemOpen
  \bibfield  {author} {\bibinfo {author} {\bibfnamefont {M.~J.}\ \bibnamefont
  {Bijlsma}}, \bibinfo {author} {\bibfnamefont {E.}~\bibnamefont {Zaremba}}, \
  and\ \bibinfo {author} {\bibfnamefont {H.~T.~C.}\ \bibnamefont {Stoof}},\
  }\bibfield  {title} {\enquote {\bibinfo {title} {Condensate growth in trapped
  bose gases},}\ }\href {\doibase 10.1103/PhysRevA.62.063609} {\bibfield
  {journal} {\bibinfo  {journal} {Phys. Rev. A}\ }\textbf {\bibinfo {volume}
  {62}},\ \bibinfo {pages} {063609} (\bibinfo {year} {2000})}\BibitemShut
  {NoStop}%
\bibitem [{\citenamefont {Snizhko}\ \emph {et~al.}(2016)\citenamefont
  {Snizhko}, \citenamefont {Isaieva}, \citenamefont {Kuriatnikov},
  \citenamefont {Bidasyuk}, \citenamefont {Vilchinskii},\ and\ \citenamefont
  {Yakimenko}}]{PhysRevA.94.063642}%
  \BibitemOpen
  \bibfield  {author} {\bibinfo {author} {\bibfnamefont {Kyrylo}\ \bibnamefont
  {Snizhko}}, \bibinfo {author} {\bibfnamefont {Karyna}\ \bibnamefont
  {Isaieva}}, \bibinfo {author} {\bibfnamefont {Yevhenii}\ \bibnamefont
  {Kuriatnikov}}, \bibinfo {author} {\bibfnamefont {Yuriy}\ \bibnamefont
  {Bidasyuk}}, \bibinfo {author} {\bibfnamefont {Stanislav}\ \bibnamefont
  {Vilchinskii}}, \ and\ \bibinfo {author} {\bibfnamefont {Alexander}\
  \bibnamefont {Yakimenko}},\ }\bibfield  {title} {\enquote {\bibinfo {title}
  {Stochastic phase slips in toroidal bose-einstein condensates},}\ }\href
  {\doibase 10.1103/PhysRevA.94.063642} {\bibfield  {journal} {\bibinfo
  {journal} {Phys. Rev. A}\ }\textbf {\bibinfo {volume} {94}},\ \bibinfo
  {pages} {063642} (\bibinfo {year} {2016})}\BibitemShut {NoStop}%
\bibitem [{\citenamefont {Kunimi}\ and\ \citenamefont
  {Danshita}(2019)}]{PhysRevA.99.043613}%
  \BibitemOpen
  \bibfield  {author} {\bibinfo {author} {\bibfnamefont {Masaya}\ \bibnamefont
  {Kunimi}}\ and\ \bibinfo {author} {\bibfnamefont {Ippei}\ \bibnamefont
  {Danshita}},\ }\bibfield  {title} {\enquote {\bibinfo {title} {Decay
  mechanisms of superflow of bose-einstein condensates in ring traps},}\ }\href
  {\doibase 10.1103/PhysRevA.99.043613} {\bibfield  {journal} {\bibinfo
  {journal} {Phys. Rev. A}\ }\textbf {\bibinfo {volume} {99}},\ \bibinfo
  {pages} {043613} (\bibinfo {year} {2019})}\BibitemShut {NoStop}%
\bibitem [{\citenamefont {Bidasyuk}\ \emph {et~al.}(2010)\citenamefont
  {Bidasyuk}, \citenamefont {Vanroose}, \citenamefont {Broeckhove},
  \citenamefont {Arickx},\ and\ \citenamefont
  {Vasilevsky}}]{PhysRevC.82.064603}%
  \BibitemOpen
  \bibfield  {author} {\bibinfo {author} {\bibfnamefont {Y.}~\bibnamefont
  {Bidasyuk}}, \bibinfo {author} {\bibfnamefont {W.}~\bibnamefont {Vanroose}},
  \bibinfo {author} {\bibfnamefont {J.}~\bibnamefont {Broeckhove}}, \bibinfo
  {author} {\bibfnamefont {F.}~\bibnamefont {Arickx}}, \ and\ \bibinfo {author}
  {\bibfnamefont {V.}~\bibnamefont {Vasilevsky}},\ }\bibfield  {title}
  {\enquote {\bibinfo {title} {Hybrid method (jm-ecs) combining the $j$-matrix
  and exterior complex scaling methods for scattering calculations},}\ }\href
  {\doibase 10.1103/PhysRevC.82.064603} {\bibfield  {journal} {\bibinfo
  {journal} {Phys. Rev. C}\ }\textbf {\bibinfo {volume} {82}},\ \bibinfo
  {pages} {064603} (\bibinfo {year} {2010})}\BibitemShut {NoStop}%
\bibitem [{\citenamefont {Bidasyuk}\ and\ \citenamefont
  {Vanroose}(2013)}]{BIDASYUK201360}%
  \BibitemOpen
  \bibfield  {author} {\bibinfo {author} {\bibfnamefont {Y.}~\bibnamefont
  {Bidasyuk}}\ and\ \bibinfo {author} {\bibfnamefont {W.}~\bibnamefont
  {Vanroose}},\ }\bibfield  {title} {\enquote {\bibinfo {title} {Improved
  convergence of scattering calculations in the oscillator representation},}\
  }\href {\doibase https://doi.org/10.1016/j.jcp.2012.09.018} {\bibfield
  {journal} {\bibinfo  {journal} {Journal of Computational Physics}\ }\textbf
  {\bibinfo {volume} {234}},\ \bibinfo {pages} {60 -- 78} (\bibinfo {year}
  {2013})}\BibitemShut {NoStop}%
\end{thebibliography}%

\end{document}